\documentclass[8pt]{article}
\usepackage{caption}
\usepackage{subcaption}

\usepackage{setspace} 
\usepackage[english]{babel}

\usepackage[letterpaper,top=2cm,bottom=2cm,left=3cm,right=3cm,marginparwidth=1.75cm]{geometry}
\usepackage[table]{xcolor}
\usepackage{xcolor,colortbl}
\definecolor{LightCyan}{rgb}{0.88,1,1}
\usepackage{amsmath}
\usepackage{graphicx}
\usepackage{caption}

\usepackage[colorlinks=true, allcolors=blue]{hyperref}
\usepackage{array}
\usepackage{comment}
\newcolumntype{P}[1]{>{\centering\arraybackslash}p{#1}}

\title{  A Bayesian Projection of the Total Fertility Rate of \\Puerto Rico:  2020-2050 \\ \small{First  Version}}
\author{Ang\'elica M. Rosario Santos*, Hernando Mattei** and Luis Pericchi Guerra*\\
*Department of Mathematics, University of Puerto Rico at Rio Piedras\\ ** 
Graduate School of Public Health, University of Puerto Rico, Medical Sciences Campus}
\date{}

\begin{document}
%TC:ignore
\maketitle

\begin{abstract}
.
The abrupt decline in the Total Fertility Rate (TFR) of Puerto Rico since 2000 makes the prospect of a sustained population decline a real possibility. From 2000 to 2021 the TFR declined from 2.1 to 0.9 children per woman, one of the lowest in the world. Population projections produced by the United States Census Bureau and the United Nations Population Division show that the island population may decline from 3.8 millions in 2000 to slightly above 2 million by 2050, a dramatic 47\% population decline in 50 years. As dire as this prospect may be, this may be an optimistic scenario. Both projections have the TFR increasing to 1.5 by 2050, but a fertility projection conducted by us show that fertility can remain much closer to 1.0 until 2050.

 Bayesian Hierarchical Probabilistic Theory  \cite{raftery2012bayesian} \cite{alkema2009probabilistic} has been used by the United Nations to incorporate a way to measure the uncertainty and to estimate the projection parameters. However, the assumption that the fertility level in countries with low fertility will eventually increase to 2.1 has been widely criticized as unrealistic and not supported by evidence. We modified the assumptions used by the United Nations considering countries with TFR similar to Puerto Rico and find that by 2050 Puerto Rico may have a TFR of 1.1 bounded by a 95\% credibility interval (0.56,1.77). This indicates that there may be a larger population decline than what current projections show.
 
 \

Indexing terms:  Total Fertility Rate (TFR), Probabilistic Model for TFR, TFR Projections.

\end{abstract}
%TC:endignore

\section{Introduction}

\subsection{Puerto Rico's Demographic Trends}

Puerto Rico is facing the real prospect of a large population decline. According to the most recent United States Census Bureau Projection, Puerto Rico’s population will decline from 3.8 millions persons in 2000 to slightly above 2 million by 2050, a dramatic 47\% population decline in 50 years. Such scenario will present a major challenge for the government’s social and economic policies. We are in the early stages of a demographic crisis that must be addressed immediately. Serious short- and long-term  implications will appear due in large part to the disproportionate population decline across all age groups. A low fertility tendency, a high pattern of migration and an increasing life expectancy has generated a complete new panorama for Puerto Rico. Population projections scenarios based on robust statistical analysis are indispensable for the development and incorporation of policies that could alleviate the population collapse.

Puerto Rico's population reached its largest size in 2004 being 3,826,878 and it has been declining since then. For 2017, 3,195,153 million of people was reported representing a decline of 16.5 \%. An important demographic indicator that contributed to this fact is fertility.  Total Fertility Rate patterns have declined substantially contributing to the reduction of the population size. It reached the replacement level (2.1) in 1998 and further decline to 1.1 in 2017 ( one of the lowest in the world).

Recently, the discussion about the trends in the population has been focused on the emigration caused by the devastating impact of Hurricanes Irma and Maria. However, the Total Fertility Rate's patterns are also worrying. The decrement about 1.4 in Total Fertility Rate from 1950 to 2017, suggests no indication or tendency to increase in the next years.  During the 1950's, the Island experienced a high level of emigration, but the population grew because births compensated for the  population loss due to migration. Now, the reduction in births has been enormous and does not compensate for the population loss due to migration. Worse still, births also do not compensate for the loss of population due to deaths and changes in the age structure.
%This situation will undoubtedly continue for the foreseeable future, at least, if no far-reaching decisions are made and implemented. We estimate that migration was responsible for about half of the population loss between 2010 and 2020.
Even in the absence of migration, the low birth rates in Puerto Rico guarantee that depopulation will not stop. In this sense, the problem of depopulation and aging is the problem of low birth rates. Emigration is  some how changeable,volatile and sensitive to public policies and economic variables, but the decline in fertility is stable and monotonous, and outside, alternatives like planned immigration should be conisidered.

\subsection{Alternative Methodology for Total Fertility Rate(TFR) Projection}

In fact, the dramatic reduction in Puerto Rico's population has been studied by different entities during the last decade \cite{abel2014causes}\cite{hinojosa2019population} \cite{burrowes2004potential} \cite{cohn2014puerto} and TFR projections have been performed by U.S. Bureau of Census and other agencies by using deterministic methods. There is now agreement that it is best to use a probabilistic approach in order to measure the level of uncertainty. Recently, researchers have incorporated Bayesian Hierarchical Probabilistic Theory in order to include the uncertainty in the estimation of the basic demographic indicators of fertility and mortality used in the projections \cite{alkema2015united} \cite{alkema2009probabilistic} \cite{raftery2014joint}. These models have three major advantages. The first one is the incorporation of a way to include uncertainty in the predictions. Raftery and his colleagues \cite{raftery2012bayesian} provided a solution to this problem allowing the exchange of information (borrowing strength in a hierarchical model) among countries based
on the assumption that the unknown quantities are drawn from a common probability distribution. Other advantage of this modelling is the possibility to estimate the parameters at the same time uncertainty is considered including a heteroscedastic error term $e_{c,t}$ in each model. The third advantage is the estimation of the parameters via the Bayesian Approach instead of fixing them by an expert opinion.

\begin{comment}

\begin{figure}
    \centering
    \caption{TFR Projections for Puerto Rico by US.Census Bureau}
    \includegraphics[scale=0.2]{chart.jpeg}
     \label{fig:my_label}
\end{figure}

\end{comment}
We studied the probabilistic Total Fertility Rate (TFR) model  \cite{raftery2012bayesian} that is now used by the United Nations Population Division to perform new projections. Since data used by the US. Census Bureau agency look overestimated (see Figure 1), we incorporate actualized data to have realistic scenarios. Contrary to the TFR projections performed by the US. Census, United Nations provides a way to include uncertainty. The TFR methodology used by United Nations is divided into three phases and it is later explained in detail. We firmly believe that the estimation of the AR(1) model proposed for Phase III (Recovery around the replacement level 2.1) has room for improvement if we take into account another condition for countries with low fertility rates. The idea of the authors is that countries that are not yet in Phase III will be influenced by countries that have seen two increments below $TFR = 2.1$. Therefore, after the estimation of a AR(1) hierarchical model with countries in Phase III, it is then applied to all countries. Our observation is that for countries like Puerto Rico, which have very low fertility levels,
the consideration of countries which are in Phase III but also has low fertility levels could provide a better way to "borrowing strength" among countries.

Therefore, we explored the use of a Bayesian Hierarchical Model only for countries with TFR less than 1.5. After analyze the scenarios, we propose the consideration of a low-fertility distribution (instead of a world distribution) for countries with TFR less than 1.5 to produce a more plausible realistic projection of Total Fertility Rate for Puerto Rico. This alternative modelling approach does not requires the same memory and computational time of the first modelling with all countries being a faster way to produce TFR plausible projections for these group of countries.

\section{Material and Methods}

\subsection{Bayesian Modelling Approach Notes}

Bayesian Probabilistic Methodology for projecting Total Fertility Rate was mainly proposed by Raftery et. al \cite{raftery2012bayesian}. In the last decade \cite{bijak2008bayesian}\cite{raftery2012bayesian} many researchers have opted for this statistical approach to perform projections, proposed new models and doing critical analysis. The mathematical background to study demographic patterns was mostly based on deterministic models being the Cohort Component Method the most used and acceptable method. However, the Cohort Component Method does not take into account statements about uncertainty in the estimations. Raftery showed that a wonderful way to account for uncertainty is to use the Bayesian Approach.
This paradigm is based on the Bayes Theorem and consists in utilize the sample information to update our prior knowledge about a random variable $\theta$ that represents a object under study into the posterior knowledge \cite{raftery2012bayesian} \cite{bijak2008bayesian}.

\begin{center}
\textbf{Bayes Theorem}
\begin{equation}
     p(\theta|y)= \frac{p(y|\theta)p(\theta)}{p(y)}
\end{equation}
\end{center}

In the last equation, $p(y|\theta)$ is the likelihood of the data, $p(\theta)$ is the prior distribution of $\theta$ and 
$p(y)$ known as the marginal distribution and indicates the quality of the model against the available data.

One of the brilliant ideas that Raftery incorporated to the deterministic models previously used by the United Nations was the implementation of Bayesian Hierarchical Models.  The main idea of Bayesian hierarchical modelling is to use the available data to "borrowing strength" and take advantage of the dependency among groups. For example, for the TFR projections, country-specific information on the expected maximum decline in a five-year period is limited for any country which has only just started its fertility decline. Raftery and his colleagues provided a solution to this problem allowing the exchange of information (borrowing strength in a hierarchical model) between countries based on the assumption that the unknown quantities are drawn from a common probability distribution. Another advantage of this modelling is the possibility to estimate the parameters at the same time uncertainty is considered including a heteroscedastic error term $e_{c,t}$ in each model. In the past, parameters were assigned by experts and the sense of uncertainty in the predictions was very related with the consideration of several scenarios for demographic components as Total fertility Rate. Each Bayesian Hierarchical Model produces a large number of possible future trajectories from the posterior predictive distribution instead of trajectories based only on past patterns.
In the next section, the proposed model for Total Fertility Rate is explained in detail.

\subsection{Bayesian Hierarchical Model for Total Fertility Rate}

The most used measure of the overall level of fertility is the Total Fertility Rate(TFR) defined as the average number of children a woman would bear in her life  if exposed to the age specific mortality rates prevalent at time t. TFR starts from a high level that differs among countries but starts to decline slowly. \cite{raftery2012bayesian}. United Nation's (UN) experts considered necessary to divide the Total Fertility Rate Model into three phases. 

\begin{itemize}
\item Phase I precedes the beginning of the fertility transition and is characterized by high fertility that is stable or increasing \cite{raftery2012bayesian}. They did not consider this phase in the model because all countries have now completed this phase.
\item Phase II consists of the fertility transition during which fertility declines from high levels to below the replacement level of 2.1 children per woman.
\item Phase III is the post fertility transition period. It starts after the fertility transition has been completed. %Specifically, the start of Phase III is in the middle of two increments below a TFR of two children \cite{raftery2012bayesian}.
   
\end{itemize}

 When a country is in Phase II, the five-year decline in its TFR is modeled as a double logistics function. The sum of the two logistic functions is a parametric function that describes a decline in fertility that starts with a slow pace at high TFR values. The resulting model for Phase II is as follows:

\begin{equation}
    { f }_{ c,t+1 }\quad ={ \quad f }_{ c,t }\quad -\quad g({ f }_{ c,t }\quad |{ \theta  }^{ c })\quad +{ \quad a }_{ c,t }
\end{equation}

where the five year decrement $g({ f }_{ c,t })$ is given by:

\begin{equation}
   \tiny{ g({ f }_{ c,t }|{ \theta  }^{ c })=\frac { { -d }^{ c } }{ 1+\quad exp(-2ln(9)({ f }_{ c,t }-\sum _{ i=2 }^{ 4 }{ { \triangle  }_{ i }^{ c } } +0.5{ \triangle }_{ 1 }^{ c })/{ \triangle  }_{ 1 }^{ c }) } +\frac { { d }^{ c } }{ 1+\quad exp(-2ln(9)({ f }_{ c,t }-{ \triangle }_{ 4 }^{ c }-0.5{ \triangle  }_{ 3 }^{ c })/{ \triangle  }_{ 3 }^{ c }) } }
\end{equation}

with ${ \theta }^{ c }=({ \triangle  }_{ 1\quad  }^{ c },{ \triangle  }_{ 2\quad  }^{ c },{ \triangle  }_{ 3\quad  }^{ c },{ \triangle  }_{ 4 }^{ c },{ \quad d }^{ c })$ being a vector  of country-specific -parameters and ${ \quad a }_{ c,t }\overset { ind }{ \sim  } N(0,\sigma (t,{ \quad f }_{ c,t })^{ 2 })$ where $\sigma$ is a function that  describes how the error standard deviation changes with fertility level and time period. The prior-distributions assigned to these parameters are explained in detail in \cite{alkema2011probabilistic}.

They define a country as having entered Phase III once, two consecutive five-year increases below a TFR of 2 children have occurred.The resulting model is as follow:

\begin{equation}
    { f }_{ c,t+1 }\quad -\mu_{c} ={ \quad \rho (f }_{ c,t }\quad -\quad \mu_{c} )\quad +{ \varepsilon }_{ c,t }
\end{equation}

where ${ \varepsilon }_{ c,t }\overset { i.i.d. }{ \sim  } N(0,{ \sigma  }_{ \varepsilon}^{ 2 })$ and \begin{equation}
{ \mu  }_{ c } \sim { TN }_{ [0,\infty ) }(\bar { \mu  } ,{ \sigma  }_{ \mu  }^{ 2 })\quad ; \quad { \rho  }_{ c } \sim { TN }_{ [0,\infty ) }(\bar { \rho  } ,{ \sigma  }_{ \rho  }^{ 2 })
\end{equation} \begin{equation}
\bar { \mu  } \sim U[0,2.1],\quad { \sigma  }_{ \mu  }\sim U[0,0.5]\quad ,\quad \bar { \rho  } \sim U[0,1]\quad ,\quad { \sigma  }_{ \rho  }\sim U[0,0.289],{ \quad \sigma  }_{ \varepsilon  }\sim U[0,0.5]
\end{equation}

\subsubsection{Annual TFR Projections}

The modelling techniques presented above were developed to work with five-year data. In \cite{liu2022probabilistic} there is an alternative to estimate and project annual data if desired.
In order to do that, {\v{S}}ev{\v{c}}{\'\i}kov{\'a}, Raftery and Liu modified the Phase II model by adding an additional first-order autorregresive component. Now the TFR decrement is modelled as:

\begin{equation}
    d_{c,t+1}-g(\theta_{c},f_{c,t+1})= \phi ( d_{c,t}-g(\theta_{c},f_{c,t})) + a_{c,t}
\end{equation}

where the prior distribution of $\phi= U(0,1)$ and the distribution of the random distortions is the same as the five-year-model. For most parameters the same distribution fixed for the five year model is used.
There is an exception for the prior distribution of $\sigma_{0}$. For a detailed explanation see \cite{liu2022probabilistic}.

When annual data is used no changes have been made to Phase III. However, to determine the start of the phase, five years averages of TFR are obtained and then the same rule is applied meaning that Phase III starts when two consecutive increases of TFR below 2.1 are observed.

\subsubsection{Alternative TFR - Projections Scenario for Low TFR countries}

A study performed in 2013 called \textit{Future Fertility in low fertility countries}\cite{basten2014future} suggests, after doing several surveys to international experts, that given the wide variety of fertility levels in rich and middle- income countries a global convergence of fertility around replacement level 2.1 appears unlikely. The authors pointed out that any understanding of the factors behind the fertility decline needs to take into account the changes, like economics, in Old Europe, Latin America, and the Middle East.

Puerto Rico is one of the countries having low TFR and this assumption should be reconsidered and taking into account. %The pace of increment(slow or fast) depends on the autorregresive parameter $\rho$ in the Phase III model described above. 
%Note that the expected TFR is calculated as $Expected\quad TFR=2.1 - (2.1-TFR)*\rho$. 
Our alternative way to produce the projections is to restrict the world distribution to countries also showing this pattern of low fertility TFR instead of all countries in WPP-2022. In this case the property of "borrowing strength" of the hierarchical modelling could be very useful to produce very justified TFR projections. The drift of the TFR random walk model given by $g({ f }_{ c,t }\quad |{ \theta  }^{ c })$ describing the decline in fertility will be now obtained via a low-fertility countries probability level. %However since all countries have experimented Phase II, the estimation of the parameters is not considerable affected.

\section{Results}

Due to the overestimation of the TFR data for Puerto Rico in the World Population Prospects 2022 report, and the assumption of a TFR=1.5 made by U.S. Census Bureau in their TFR projections, we consider two approaches for producing plausible TFR Projections by using bayesTFR package in $\mathbf{R}$ 

\subsection{Two Modelling Approaches for TFR Projections for Puerto Rico}

\begin{figure} [h!]
     \centering
     \begin{subfigure}[b]{0.3\textwidth}
         \centering
         \includegraphics[width=\textwidth]{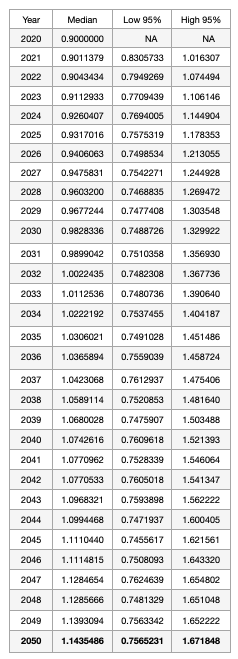}
         \caption{Modelling 1:Total Fertility Rate Projections
         for Puerto Rico for Hierarchical Model 
         with all countries}
     \end{subfigure}
     \hfill
     \begin{subfigure}[b]{0.305\textwidth}
         \centering
         \includegraphics[width=\textwidth]{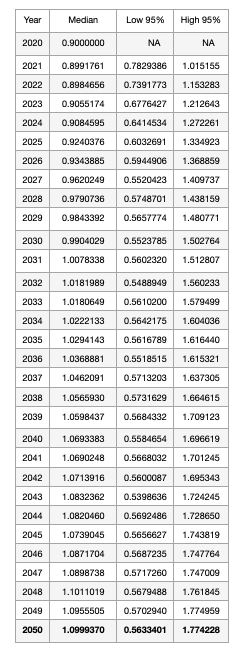}
         \caption{Modelling 2: Total Fertility Rate Projections 
         for Puerto Rico for Hierarchical Model with 
         17 countries with low TFR values}
     \end{subfigure}
     \hfill
        \caption{Two Modelling Approaches for Puerto Rico's TFR Projections}
     \label{Projections}   
\end{figure}

The first approach take into consideration the data for all countries for the fitting of a  Bayesian Hierarchical Model as described by Raftery and colleagues. The second scenario is based on segregating a set of countries characterized by low fertility rates below 1.5. It sounds reasonable since it would permit that parameters’ estimates for low fertility countries be only affected by the low fertility countries -level probability instead of by world-level probability.

\subsection{Cross Validation : TFR Projections for Puerto Rico since 2000}
To test our models we perform a cross validation scenario starting the projections at 2000 . This permit us to observe how good the models predict the next twenty years. TFR projections for Puerto Rico are showed in Figures below. For each TFR projection we get a 95\% credibility interval and a set of estimated parameters from the posterior distribution. 

\begin{figure}[h!]
    \centering
   \includegraphics[width=0.7\textwidth]{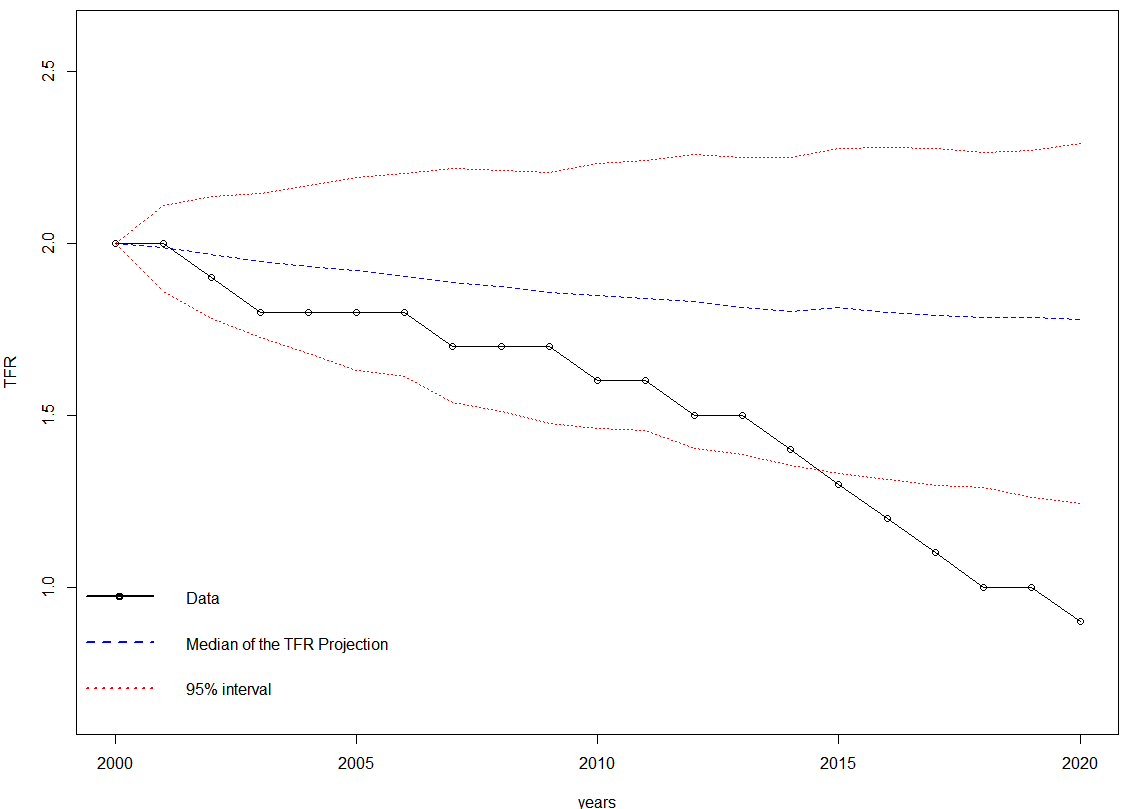}
    \caption{Total Fertility Rate Projections for Puerto Rico since 2000
     for Hierarchical Model with all countries}
    \label{fig:my_label}
\end{figure}

\

\begin{figure}[h!]
    \centering
   \includegraphics[width=0.7\textwidth]{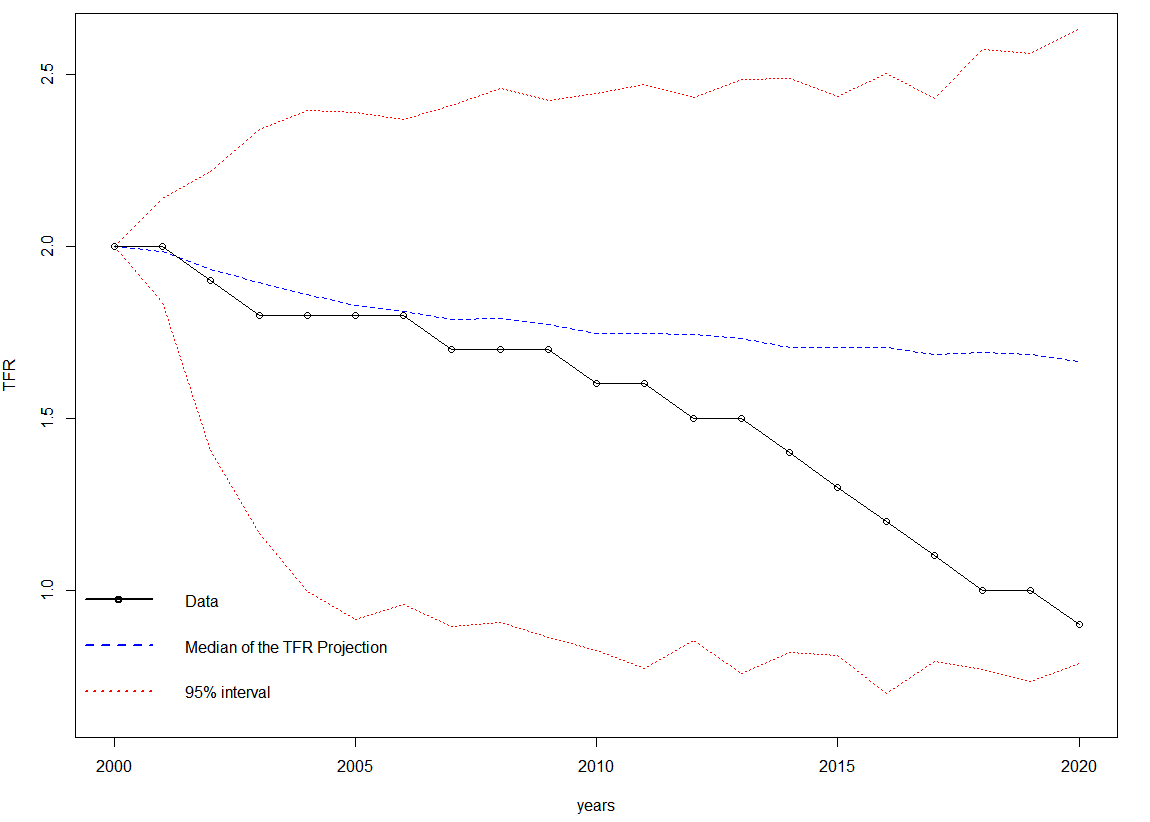}
    \caption{Total Fertility Rate Projections for Puerto Rico since 2000
     for Hierarchical Model with 17 countries with low TFR values}
    \label{fig:my_label1}
\end{figure}

\

\newpage

\subsubsection{Goodness of fit of the Double Logistic Function}

\

For the two modelling approaches characterized by the amount of countries, it is important to consider and compare some measures of goodnees of fit. Coverage is the ratio of observed data fitted within the 95\% probability interval of the predictive posterior distribution of the double logistic function. In the tables below are other measures of fit like the root mean square error and mean absolute error of the simulation for the TFR data from 1950-2000 (Cross Validation data) and 1950-2020 (Data for TFR projections). Total Coverage is the coverage for the TFR data for all the considered countries in each modelling approach.

\begin{table}[h!]
\scriptsize
\centering
\begin{tabular}{ |p{2.1cm}|P{3.0cm}|P{3.0cm}|}
\hline
\rowcolor{lightgray} \multicolumn{3}{|c|}{\textbf{Goodness of fit of the double logistic function (1950-2000)}} \\
\hline
\textbf{Measure of fit }&Modelling 1& Modelling 2 \\
\hline
1) Total Coverage 95\% &95\%& 94\% \\
\hline
2) Total Root Mean Square Error (RMSE) &0.0873&0.1433\\
\hline
3) Total Mean Absolute Error (MAE) & 0.0456&0.0735\\
\hline
3) Puerto Rico 95\% Coverage &88\%&98\%\\
\hline
4) Republic of Korea 95\% Coverage &98\%&100\%\\
\hline
5)Cuba 95\% Coverage &84\%&88\%\\
\hline
\end{tabular}
\caption{Goodness of fit of the double logistic function(1950-2000)}
\label{table:Tabla 2}
\end{table}

\begin{table}[h!]
\scriptsize
\centering
\begin{tabular}{ |p{2.1cm}|P{3.0cm}|P{3.0cm}|}
\hline
\rowcolor{lightgray} \multicolumn{3}{|c|}{\textbf{Goodness of fit of the double logistic function(1950-2020)}} \\
\hline
\textbf{Measure of fit }&Modelling 1& Modelling 2 \\
\hline
1) Total Coverage 95\% &94\%& 94\% \\
\hline
2) Total Root Mean Square Error (RMSE) &0.0783&0.1312\\
\hline
3) Total Mean Absolute Error (MAE) & 0.0420&0.0687\\
\hline
3) Puerto Rico 95\% Coverage &88\%&98\%\\
\hline
4) Republic of Korea 95\% Coverage &97\%&100\%\\
\hline
5)Cuba 95\% Coverage &76\%&90\%\\
\hline
\end{tabular}
\caption{Goodness of fit of the double logistic function(1950-2020)}
\label{table:Tabla 3}
\end{table}

\section{Discussion}

We have considered two modelling approaches for Total Fertility Rate as shown in Figure \ref{Projections}. The convergence of the MCMC (Markov Chains Monte Carlo) chains was adequately checked for each generated modelling. Both models coincide in a TFR=1.1 for 2050, a value not too far from the actual TFR=0.9.
However, some remarkable differences between them will help us to understand why the second modelling is an acceptable alternative way to produce TFR Projections for Puerto Rico. Contrary to the first modelling in which the data of all countries is taken into consideration, the second modelling selects countries according to the criteria of a  $TFR\leq 1.5$, providing a faster and computationally much better approach to reach the same objective. To compare the modellings' performance we test a kind of cross-validation procedure by performing TFR projections for each one since 2000. In this case, is important to point out
that Phase III for both modellings is estimated by using fewer countries since for 2020, as data arrives, there will be more countries entering this Phase. 

Results are shown in Figures \ref{fig:my_label} and \ref{fig:my_label1}. From Figure \ref{fig:my_label} is evident that the 95\% credibility interval of the TFR projection in Modelling 1 contains the data registered for Puerto Rico only until 2014. After this year, TFR values fall outside the 95\% credibility interval. The tendency of the dramatic TFR decline for Puerto Rico is atypical and of course, when actualized data is added to the model the median and the credibility intervals are automatically adjusted. Now let's see Modelling 2. The main difference with the first scenario is the width of the credibility intervals. This modelling considers much lower values for the low 95\% credibility band. Therefore, all the TFR data registered for two periods(2001-2010, 2011-2020) are contained in the 95\% credibility interval. Clearly, for these two periods, the TFR descent starting from the replacement level 2.1 to low TFR values was better captured by the second modelling. In summary,  Figures \ref{fig:my_label} and \ref{fig:my_label1} points out that in the last twenty years Modelling 2 has clearly been closer to the real data than Modelling 1.

For data since 1950-2000 and 1950-2020, Tables \ref{table:Tabla 2} and \ref{table:Tabla 3} show the Goodness of Fit of the Double Logistic function. Observe that the total coverage is similar. But the 95\% coverage for countries as Puerto Rico, the Republic of Korea, and Cuba show a significant improvement supporting Modelling 2. Although we have little more MAE and RMSE , the ratio of observed data fitted within the 95\% probability interval of the predictive posterior distribution of the double logistic function is considerable much better.

\section{Conclusion}

Overall, the main goal of our work was achieved. We produce an alternative way to perform TFR projections for countries with low TFR such as Puerto Rico. Different to US. Census and United Nations Projections, we expect a TFR=1.1(0.56,1.77) by 2050.

Observe that our projection still considers the possibility of a TFR=1.5 by 2050 but not such as the most plausible scenario. It is important to say that changes in fertility are very associated to social conducts, and implementation of public policies could alleviate in some way the rapid and continuous drop in births. This possibility is considered in our projections up to the upper level credibility band.

Further research in determinants of fertility is needed in order to strengthen our modelling and understand the social behavior related to child-bearing. The latest study of fertility determinants in Puerto Rico was performed in 1987
\cite{warren1987fertility}, and it is urgent to perform a new study.

\section{Acknowledgements}
This research project was supported by Brigde To the Doctorate Program Cohort XIII, Grant Number: HRD-1906130 and DEGI Thesis/Dissertation Scholarship granted by Univeristy of Puerto Rico at Rio Piedras.

\bibliographystyle{ieeetr}
\bibliography{sample}
\end{document}